\documentclass[conference]{IEEEtran}
\IEEEoverridecommandlockouts
\usepackage{amsmath,amssymb,amsfonts,amsthm}

\usepackage{xcolor}
\usepackage{algorithmic}
\usepackage{graphicx}
\usepackage{textcomp}
\usepackage{url}
\usepackage{pgfplots}
\usepgfplotslibrary{groupplots}
\pgfplotsset{compat=1.16}
\usepackage{caption}
\usepackage{subfig}
\usepackage{algorithm}
\usepackage{booktabs}
\usepackage{array} 
\usepackage{balance}
\usetikzlibrary{calc,tikzmark}

\pgfdeclarelayer{background}
\pgfsetlayers{background,main}
\newcommand*\circled[1]{\tikz[baseline=(char.base)]{
            \node[shape=circle,draw,inner sep=0.5pt] (char) {#1};}}

\def\BibTeX{{\rm B\kern-.05em{\sc i\kern-.025em b}\kern-.08em
    T\kern-.1667em\lower.7ex\hbox{E}\kern-.125emX}}

\DeclareMathOperator{\mean}{\mathbb{E}}
\newcommand{\be}{\begin{equation}}
\newcommand{\ee}{\end{equation}}

\newcommand{\Bd}{\hbox{Beta}}
\newcommand{\kmv}{\textsf{KMV}}

\newtheorem{thm}{Theorem}

\begin{document}

\title{Key Compression Limits for $k$-Minimum Value Sketches
}

\author{\IEEEauthorblockN{Charlie Dickens}
\IEEEauthorblockA{\textit{Yahoo!}}
\and
\IEEEauthorblockN{Eric Bax}
\IEEEauthorblockA{\textit{Yahoo!}}
\and
\IEEEauthorblockN{Alexander Saydakov}
\IEEEauthorblockA{\textit{Yahoo!}}
}

\maketitle

\begin{abstract}
The $k$-Minimum Values (\kmv) data sketch algorithm stores the $k$ least hash keys generated by hashing the items in a dataset. 
  We show that compression based on ordering the keys and encoding successive differences can offer $O(\log n)$ bits per key in expected storage savings, 
  where $n$ is the number of unique values in the data set. 
  We also show that $O(\log n)$ expected bits saved per key is optimal for any form of compression for the $k$ least of $n$ random values 
  -- that the encoding method is near-optimal among all methods to encode a \kmv sketch.
  
  We present a practical method to perform that compression, 
  show that it is computationally efficient, and demonstrate that its average savings in practice is within about five percent of the 
  theoretical minimum based on entropy. 
  We verify that our method outperforms off-the-shelf compression methods, 
  and we demonstrate that it is practical, using real and synthetic data.
\end{abstract}


\section{Introduction}

Big data analysis is extremely resource-intensive, even for seemingly simple queries such 
as counting unique items in a dataset.
Sketching reduces the computational burden \cite{cormode2020small}, however, it does 
not make computation free.
One application that particularly benefits from sketching algorithms is reporting tools that
process billions of events per day for large technology companies \cite{bailis2017macrobase}.
Such tools are practical because they exploit the mergeability of sketches: a sketch can be 
generated for a fine-grained time interval, for example hourly, and then hourly sketches
can be merged into a daily sketch, and daily sketches can, in turn, be merged into a weekly sketch and so on.
Repeating this process to generate reports for a three month rolling window, for example, requires writing sketches to disk after processing their respective data streams, then merging thousands of sketches at a time.

Although individual sketches are small, the aggregate space consumed by sketches imposes a noticeable cost on enterprises.
Storage \emph{is not free} \cite{cloud_storage}, and maintaining millions of sketches is expensive. 
We are interested in compressing data sketches as efficiently as possible -- without requiring prohibitive amounts of computation -- to reduce storage while improving scalability of analysis.

Historically, research on sketching algorithms has focused on quantities such as the 
accuracy, (asymptotic) space usage and update time (see \cite{kane2010optimal} for a full review).
Early attempts at sketch compression focused on \emph{Probabilistic Counting with Stochastic Averaging} (PCSA), exploiting the statistical structure of sketches stored as bitmaps \cite{flajolet1985probabilistic}.
Despite this early effort, compressibility of distinct counting sketches largely fell out of focus in the research literature until recently.
Instead, efforts were made to optimize algorithmic efficiency.
This approach resulted in the LogLog, SuperLogLog, and HyperLogLog algorithms \cite{durand2003loglog,flajolet2007hyperloglog} which reduce the 
sketch size by using $O(\log \log n)$ bits per bucket rather than the $O(\log n)$ bits required for PCSA.

These approaches reduced the algorithmic space complexity but did not address compressibility of sketches.
However, compressed sketches are vitally important in practice: smaller sketches consume less space, so they cost enterprises 
less to store, and smaller objects are energy-efficient and quicker to send across a distributed network.
This means that a reduction in the size of the stored sketch can have a substantial benefit on performance in large-scale distributed systems.
Consequently, there has been a recent drive to compress sketches as efficiently as possible,
measured in terms of how close the expected number of bits in the compressed sketches are to the entropy 
of the underlying random process defining the sketching algorithm \cite{pettie2021information,pettie2020non,lang2017back,karppa2022hyperlogloglog}, because the entropy is a lower bound on expected sketch size, by Shannon's Source Coding Theorem \cite{shannon,mackay03,cover06}. 

We study the compressibility of $k$-minimum values (\kmv) sketches \cite{bar2002counting}. 
This sketch is a general framework that uses a hash function
to deduplicate the entries in a data set and stores the least $k$ hash function outputs over the entries.
This algorithm is also known as MinCount \cite{giroire2009order}, Theta sketch \cite{dasgupta2015framework},
or bottom-$k$ \cite{cohen2007summarizing}.
The \kmv{} class of sketches is particularly attractive because they are more flexible than 
other distinct counting sketches --
they permit unique count queries over the union of data streams, as well as arbitrary set operations \cite{dasgupta2015framework}.
This produces lower error than the inclusion-exclusion rules for set operations used with HLL and other cardinality estimators \cite{dasgupta2015framework}.
For some intuition and mathematical background on why single-hash methods like \kmv{} perform 
well for set intersection over many sets see \cite{pagh2014min}.

\textbf{Related Work.}
There is a plethora of sketching work for distinct counting algorithms beginning with 
\cite{flajolet1985probabilistic}, which proposed the \emph{Probabilistic Counting with Stochastic Averaging} algorithm.
PCSA maintains a $k \times B$ bitmap that is initially all zeros.
On observing an item, two hash keys are generated: the first selects a row uniformly at random from $[k]$, and the second determines the column $j$ in that row to set to one, with $j$ the number of leading zeros in the second hash key.
Instead of keeping the full array, the sketch can store only the highest-order one-valued index $j$ in each row and still support cardinality estimation. 
This observation led to the HyperLogLog (HLL) algorithm \cite{flajolet2007hyperloglog}.
HLL has seen impressive practical uptake, being deployed in Google Bigquery \cite{bigquery-hll,heule2013hyperloglog} and Apache Software Foundation DataSketches \cite{asf-datasketches}.
Algorithms such as \kmv{} \cite{giroire2009order,bar2002counting,dasgupta2015framework} rely
on order statistics of keys from the hash function and have less favorable space-error tradeoffs for counting set unions compared to HLL.
However, \kmv{} sketches can be modified to estimate arbitrary set expressions, such as intersections and set differences, over the data
more accurately than HLL \cite{dasgupta2015framework}.

Some space-optimal sketches are described in \cite{kane2010optimal,blasiok2019optimal}. 
They require substantial computation and are not known to be practical.
In terms of compressed cardinality estimators, we are only aware of the \emph{Compressed Probabilistic Counting} 
(CPC) algorithm by Lang \cite{lang2017back}, the compressed LogLog algorithms of \cite{durand2004combinatoire}, and 
compressed HyperLogLog \cite{karppa2022hyperlogloglog}.
The CPC algorithm compresses the PCSA sketch (which has $4.7$ bits of entropy per row) to within $10\%$ of its entropy using off-the-shelf compression such as zstd, but  
a more sophisticated sliding window with exception handling can compress it to within $0.2$ bits of the entropy \cite{lang2017back}.
It offers the best space-accuracy tradeoff among distinct counting sketches deployed in the Apache DataSketches Library.
The LogLog algorithm is known to have about $3$ bits of entropy per bucket but uses a suboptimal estimator so is no longer favored in practice \cite[p. 136]{durand2004combinatoire}.
In comparison, \cite{karppa2022hyperlogloglog} shows how to compress the HLL algorithm to $3$ bits per bucket while retaining the 
accuracy of HLL estimation.
Compressed encodings of HLL have seen industry deployment, with the $4$-bit HLL in \cite{asf-datasketches} being the default setting.
For a theoretical perspective on sketch compression, refer to \cite{pettie2021information}. 
Smaller sketches can be obtained if we do not require mergeability of sketches \cite{pettie2020non}, but in large-scale distributed systems for industry, this is not an acceptable compromise.

Qualitatively, our results differ from both \cite{lang2017back} and \cite{karppa2022hyperlogloglog} in two key ways.
First, we study the more flexible \kmv{} sketch family which naturally admits accurate approximations of distinct count and arbitrary set operations 
rather than only distinct counts.
Second, in contrast to the distinct counting algorithms above, we do not change the internal update procedure of the algorithm, so compression can simply be added as post-processing to any existing implementation. 
Our compression is lossless so all estimation results and performance associated with a \kmv{} implementation are maintained.
We demonstrate this empirically by adding compression functionality to the open source DataSketches \kmv{} implementation.

Our focus is on compressing \kmv{} sketches, which has limited prior work.
However, there is prior work on compressing ``minwise'' hashing sketches, which are similar to \kmv{} sketches. 
But that work yields summaries that are not mergeable \cite{li2011theory} or use lossy compression \cite{yu2020hyperminhash}.
A heuristic compression scheme is given for \kmv{} sketches in \cite{harris2021distributed} in the distributed setting;
the communicating parties compute a threshold that represents the fraction of their keys each party should send over the network.
Each sketch is compressed by sending only a sample of its smallest hash keys that is determined by the cardinality estimate of 
a single sketch compared to the aggregate of the union estimate.
This is not comparable to our work because the compression of a single sketch depends on the cardinality of all other sketches.
This makes it impractical in a large, distributed system, in which the sketches being merged may have been written to disk months or years ago.
Applying these ideas to our motivating example (a streaming reporting system) would require writing all sketches to disk in full size, 
then reading them back into memory to recompress them on a subsequent write operation.
Moreoever, \cite{harris2021distributed} shows that performance degrades with the number of merges;
after only one hundred merges the relative accuracy of a union estimate decreases by more than $10\%$.
This removes a huge benefit of distinct counting algorithms, such as the \kmv{:} that they are 
fully mergeable -- accuracy does not degrade as the number of merge operations increases.
In summary, to the best of our knowledge, our results are the first to show a theoretical guarantee on the compressibility of \kmv{} sketches that 
is simultaneously (i) lossless; (ii) retains the error bounds from the original sketch for all set operations and (iii) permits full mergeability of multiple sketches 
without any loss of accuracy.

\textbf{Contributions.}
To the best of our knowledge, we present the first work on compressing \kmv{} sketches, and we develop an 
implementation that achieves near-optimal compression rates and is computationally efficient. 
Our scheme saves an expected $O(\log n )$ bits per hash key, and we show that this is optimal for compressing the least $k$ of $n$ hash keys.
This work is motivated by the consistent suboptimal performance of off-the-shelf compressors, as reported in our practical findings;
a brief example can be seen in Table \ref{tab:basic_performance}.
We empirically validate our theoretical results by adding compression functionality to the open source DataSketches \kmv{} implementation.
Our method is robust across all input cardinalities, always returning a compressed sketch, unlike generic compression methods which
often inflate the sketch size.
In the large $n$ regime, our method is within about $5\%$ of the entropy bound, so it is extremely near the limit at which 
no further improvements can be made.
Our method is a simple, time-efficient post-processing step that can be added to any existing \kmv{} implementation so we retain all
the functionality of \kmv{} without any change to error bounds or practical performance.

\begin{table}[hb]
    \caption{Compression Summary.
    The ``Compression'' column evaluates the ratio between sizes (in bytes) of a compressed sketch and an uncompressed sketch (see \eqref{eq:comp-defn}).
    Entropy ratio is the ratio between sizes (in bytes) of a compressed sketch and the entropy of the sketch.
    Instance size is $n=8.4 \times 10^{6}$.}
    \label{tab:basic_performance}
    \centering
    \begin{tabular}{@{}rrrr@{}} \toprule
      Method      & Size (bytes)     & Compression & Entropy Ratio \\ \midrule
      Uncompressed& 32792            & $-$         & $1.55$\\
      bzip2       & 30727            & 0.94        & $1.45 $\\
      zlib        & 28064            & 0.86        & 1.32  \\
      \textbf{Our method} & \textbf{22418} & \textbf{0.68} & \textbf{1.06} \\
      \hline
      Entropy bound & 21219       & 0.65             & $-$   \\
      \bottomrule
      \end{tabular}
\end{table}

\newpage
\textbf{Paper Outline.}
This paper has two parts: Sections \ref{notation_sec} and \ref{sec:experiments} introduce and evaluate the compression scheme from a practical and experimental point of view, then Sections \ref{hbs_sec} 
onwards develops entropy bound that show our compression method is near-optimal.
\textbf{Section \ref{notation_sec}} describes the \kmv sketch and introduces notation that we use in the rest of the paper. 
\textbf{Section \ref{sec:experiments}} discusses the implementation and gives empirical results for it, 
showing that the compression is within about 5\% of the limits indicated by entropy.
\textbf{Section \ref{hbs_sec}} shows that the entropy of the distribution of successive differences between (ordered) hash keys
(as unsigned integers) is approximately $\ln s - \ln n + \frac{n - 1}{n}$, where $s$ is the size of the space of hash keys: 
$s = 2^b$ where $b$ is the length of a hash key in bits. 
\textbf{Section \ref{leading_zero_sec}} shows that the minimum number of leading zeros over all successive differences is 
about $\log_2 n - \log_2 \ln n - 2$, with high probability.
So we can hope to save $O(\log_2 n)$ bits per key by removing leading zeros and encoding successive differences, on average. 
\textbf{Section \ref{hkn_sec}} estimates the entropy of the distribution of the least $k$ of $n$ random samples from a 
size-$s$ set to be $k[\ln s - \ln (n + 1) +1]$. 
That entropy limits the expected compression rate for \textit{any} scheme to encode the \kmv sketch, 
so it indicates that the best possible compression is on the order of  $\ln n$ savings per key. 
\textbf{Section \ref{discussion_sec}} concludes with some directions for future work. 
\textbf{Appendix \ref{sec:deferred-proofs}} presents proofs for most of our results and 
\textbf{Appendix \ref{emp_mean_sec}} uses empirical means to show that the approximations in 
Sections \ref{hbs_sec} and \ref{hkn_sec} are accurate to within a small fraction of a bit per key.

\section{Sketch, Assumptions, and Notation} \label{notation_sec}
In a data set that is a sequence of entries, let $n$ be the number of distinct entries and let $\mathcal{D}$ be the set of distinct entries. 
A \kmv sketch applies a hash function to each entry, producing a hash key with $b$ bits. 
For purposes of analysis, we assume that the hash function produces a hash key uniformly at random among the $s = 2^b$ length-$b$ bit vectors;
that it produces the same output for the same input, and also that $n$ is sufficiently smaller than $s$ so that the probability of two different entries having
the same hash function output is small enough that we can ignore it in our analysis. Thus we assume that the \kmv sketch produces $n$ \emph{distinct} hash keys. 
Then, treating each hash key as an unsigned integer with high-value to low-value bits from left to right, the sketch keeps the $k$ least hash keys and discards the others. (Some \kmv sketch algorithms keep the least $k$ \cite{bar2002counting}; others keep the least $k + 1$ \cite{cohen2009leveraging,dasgupta2015framework} to achieve unbiased estimators. For those, substitute $k + 1$ for $k$ in the results from this paper.)

In this paper, we evaluate the entropy of some methods of communicating the sketch in order to estimate and/or bound how much compression is possible, by appealing to Shannon's source coding theorem \cite{shannon,mackay03,cover06}, 
which states that the logarithm base-two entropy of a distribution over outcomes is the minimum number of expected bits per symbol that can be achieved in compressing a message consisting of symbols drawn i.i.d. from the distribution. 
In this paper, we often use natural logarithms so that the entropy is measured in nats, but it is simple to convert the entropy to bits in base-two logarithms -- simply divide by $\ln 2$. 

We also think of the hash keys as having values in $\left[0, 1\right)$, with the value for a hash key its $b$ bits taken as an unsigned integer with the traditional encoding, and divided by $s$. Some examples: the hash code with all zero bits has value zero, the hash code with a leading one and then all zeros has value $\frac{1}{2}$, and the hash code with all ones has value $\frac{s - 1}{s}$. 
Assuming $n \ll s$, the $n$ hash codes values have a distribution that is similar to a discretized version of $n$ i.i.d. samples from the uniform distribution over $[0, 1]$, with the value being the greatest $\frac{i}{s}$ that is at most the continuous sample value, over integers $i$. In practice, $b=64$ for industrial deployments of sketching algorithms \cite{datasketches_hash};
which rely on the $64$-bit MurmurHash algorithm \cite{murmurhash}. So even if $n$ is in the billions or trillions, $n \ll s$. 

The intuition behind our approach is as follows.
The \kmv{sketch} can be viewed as keeping a sorted list of the smallest $k$ values from $n$ generated hash keys.
If the $n$ hash values are assumed to be independent and identically drawn from a uniform distribution over all possible hash values,
then the $k$ smallest hash values are the first to $k$th order statistics for a size-$n$ uniform random sample.
If we map the space of all hash values to $[0,1]$ in the obvious way,
then, for each of the $j$ in $1, \ldots, k$, the expected value of the $j$th order statistic is $\frac{j}{n+1}$. So, in expectation, each of the order statistics moves towards zero as $n$ increases.
Now, if all of the $k$ smallest hash values move closer to zero, the differences between them should decrease, in general.
When storing the differences between hash values, if these hash values are smaller, then there are more leading zeros in the values.
Consequently, there is more redundancy in the stored bit strings, so we can trim the superfluous leading zeros from the differences and 
write the remaining bits to memory to fully describe the sketch. 
Figure \ref{fig:compression_example} shows an example of performing this compression.

\section{Implementation and Empirical Performance} \label{sec:experiments}
Before presenting the compression algorithm and its performance in detail, we give a brief preview of our main theoretical result, since we use it to evaluate the algorithm. The result is Theorem \ref{hkn_thm} (proven later), which states that if we apply a hash function that returns $b$-bit outputs uniformly at random over a space of $s = 2^b$ hash codes to $n$ items and select the $k$ smallest of the $n$ hash codes, the entropy of the distribution of those $k$ hash codes is 
\begin{equation}
    H_{K|N} \approx k [\ln s - \ln (n + 1) +1].
    \tag{$\dagger$}
    \label{eq:main_result}
\end{equation}
\begin{figure*}
    \begin{tikzpicture}
        \node (table) {\begin{tabular*}{\linewidth}{@{}llllllllllllllllllllllllllllllllllllllllllllllllllll@{}}
    \cmidrule(r){1-11} \cmidrule(lr){14-24}
     \textbf{Hashes}& \multicolumn{10}{l}{Bit index}                       &  &  & \textbf{Differences}    & \multicolumn{10}{l}{Bit index}                   &  &  &  &  &  &  &  &  &  &  &  &  &  &  &  &  &  &  &  &  &  &  \\ \cmidrule(r){1-11} \cmidrule(lr){14-24}
               & 63 & $\dots$ & 7 & 6 & 5 & 4 & 3 & 2 & 1 & 0 &  &  &  & 63 & $\dots$               & 7 & 6 & 5 & 4 & 3 & 2 & 1 & 0 &  &  &  &  &  &  &  &  &  &  &  &  &  &  &  \\ \cmidrule(lr){2-11} \cmidrule(lr){15-24}
    $h[k-1]<1$ &  0 & ---     & 1 & 0 & 0 & 1 & 0 & 0 & 0 & 1 &  &  & $\Delta[k-1]$  & 0  & ---     & 0 & 0 & 0 & 1 & 0 & 0 & 0 & 1 &  &  &  &  &  &  &  &  &  &  &  &  &  &  &  &  \\
    $h[k-2]$   &  0 & ---     & 1 & 0 & 0 & 0 & 0 & 0 & 0 & 0 &  &  & $\Delta[k-2]$  & 0  & ---     &   &   &   &   &   &   &   &   &  &  &  &  &  &  &  &  &  &  &  &  &  &  &  &  \\
    $\vdots$   &    & $\vdots$&   &   &   &   &   &   &   &   &  &  & $\vdots$       &    &         &   &   &   &   &   &   &   &   &  &  &  &  &  &  &  &  &  &  &  &  &  &  \\
    $h[2]$     &  0 & ---     & 0 & 1 & 1 & 1 & 0 & 0 & 0 & 0 &  &  & $\Delta[2]$    & 0  & ---     & 0 & \textbf{1} & 0 & 0 & 1 & 1 & 1 & 1 &  &  &  &  &  &  &  &  &  &  &  &  &  &  &  \\
    $h[1]$     &  0 & ---     & 0 & 0 & 1 & 0 & 0 & 0 & 0 & 1 &  &  & $\Delta[1]$    & 0  & ---     & 0 & 0 & 0 & 1 & 1 & 1 & 0 & 1 &  &  &  &  &  &  &  &  &  &  &  &  &  &  &  \\
    $h[0] > 0$ &  0 & ---     & 0 & 0 & 0 & 0 & 0 & 1 & 0 & 0 &  &  & $\Delta[0]$    & 0  & ---     & 0 & 0 & 0 & 0 & 0 & 1 & 0 & 0 &  &  &  &  &  &  &  &  &  &  &  &  &  &  &  \\ \cmidrule(r){1-11} \cmidrule(lr){14-24}
               &    &         &   &   &   &   &   &   &   &   &  &  & $\bigvee_j \Delta[j]$& 0&---  & 0 & \underline{\textbf{1}} & 0 & 0 & 1 & 1 & 1 & 1 &  &  &  &  &  &  &  &  &  &  &  &  &  &  \\ \cmidrule(lr){14-24}
\end{tabular*}};
        \coordinate (compBottomLeft) at (4.75, -1.65);
        \coordinate (compUpperRight) at (8.75,1.125);
        \coordinate (cutBottomLeft) at (2.45,-1.65);
        \coordinate (cutUpperRight) at (4.625,1.125);
        \draw [magenta,ultra thick,rounded corners] (compBottomLeft) rectangle (compUpperRight);
        \draw [cyan, densely dashed, ultra thick, rounded corners] (cutBottomLeft) rectangle (cutUpperRight);

       \draw[thick,->] (-0.625,-1.375) -- node[below=0.05pt] {\circled{\small{A}}} (0.25,-1.375) node[below=10pt] {\circled{\small{B}}}  ;

       \coordinate (h1Right) at (-0.625,-0.975);
       \coordinate (h2Right) at (-0.625,-0.575);
       \coordinate (d2Left) at (0.25,-0.575);
       \draw[thick] (h1Right)
       -- (h2Right) 
        -- (d2Left) 
        -- cycle;

        \coordinate (hk1Right) at (-0.625,0.85);
        \coordinate (hk2Right) at (-0.625,0.45);
        \coordinate (dk1Left) at (0.25, 0.85);
        \draw[thick] (hk1Right) 
        -- (hk2Right)
        -- (dk1Left) 
        -- cycle;

        \coordinate (C) at (6.425, 0.075);
        \draw (C) circle (0.cm) node {\circled{\small{C}}};
    \end{tikzpicture}
    \caption{Illustration of the compression algorithm.
    In the ``Hashes'' table, the raw hashes are sorted: the leading one, 
    written in bold $\textbf{1}$, is further to the right for smaller hash values.
    In Step A, we set $\Delta[0] = h[0]$ and evaluate differences between the sorted hash values.
    In Step B, we find the largest number of leading zeros in the $\Delta[j]$
    which are in general not sorted; this is done by successively 
    taking the $\textsf{OR}$ of each newly computed $\Delta[j] = h[j] - h[j-1]$ and is a number between 
    $0$ and $63$ so can be stored in $6$ bits.
    In this example, it is in column $6$ and all trailing ones in the 
    bottom row are ignored as we just need the number of leading zeros.
    Finally, in Step C we set $x = 6$ and chop off the leading bits up to index 
    $6$ (in the lefthand blue dashed-line rectangle), 
    and write only the lower $x$ bits (the solid-line red rectangle).
    }
    \label{fig:compression_example}
\end{figure*}

\subsection{Implementing Compression}
We provide a C++ implementation of the encoding scheme described below over $64$-bit hash keys.
It is designed as a post-processing step for any implementation of a \kmv{} sketch and assumes only that 
the $k$ smallest hash values are presented in sorted order.
This means that the algorithm could easily be integrated into production-quality libraries such as 
Apache Software Foundation DataSketches Library \cite{asf-datasketches}.
It is highly performant: it yields encodings that are close 
to the estimated entropy, and it is fast enough to be used in large-scale distributed systems.

\textbf{Compression Algorithm Setup.}
We assume that input is an array of length $k$ whose entries are sorted such that 
$0 < h[0] < h[1] < \dots < h[k-1]$.
Each stored hash value $h[i]$ is $64$ bits long and we assume that $h[i] \ne h[i']$ for $i \ne i'$ and $h[i] \ne 0$ for all $i$.
The compression algorithm we present requires two passes through the sketch.
The first pass of the algorithm is to compute differences between adjacent values
in the sketch which are the values $\Delta_j = h[j] - h[j-1]$ so that $\Delta_0 = h[0]$ and 
$\Delta_1 = h[1] - h[0]$ and so on.
Since the values $h[i]$ are distinct, $\Delta_j > 0$ for all $j$.
The Apache DataSketches Library has a design choice to use only $b=63$ bits of the hash function\footnote{
    Hence our calculations from Theorems \ref{hbs_thm} and \ref{hkn_thm} set $s=2^{63}$ in the plots.} 
to ensure 
cross-language compatibility between C++ and Java, so the leading bit is zero for all stored hash values.
Consequently, the largest number of leading zeros in any $\Delta_j$ is 63;
thus we can write the minimum number of leading zeros in six bits.
The compression scheme calculates the values $(\Delta_j)_{j=0}^{k-1}$ and writes a compressed version of them to memory.

\textbf{Compression Algorithm.}
A first pass computes the minimum number of leading zeros, $m$, over all $(\Delta_j)_{j=0}^{k-1}$.
To do this quickly, we $\textsf{OR}$ each successive $\Delta_j$ and compute the minimum 
number of leading zeros only once at the end of the pass.

Then we chop off the $m$ leading bits from every $\Delta_j$, leaving $k$
values $(\delta_j)_{j=0}^{k-1}$ which are just the values $(\Delta_j)_{j=0}^{k-1}$ in the final $x = 64 - m$ bits.
This has the practical advantage that all remaining values $(\delta_j)_{j=0}^{k-1}$ have fewer than 
$64$ bits but also each $\delta_j$ has the same number of bits.
Hence, we can call a single tailored subroutine $\textsf{pack}_x(\cdot)$ that specifically packs $x$ bits at a time as efficiently as possible.
The function $\textsf{pack}_x(\cdot)$ takes as input eight $\delta_j$ values and produces exactly $x$ bytes to represent these 
$\delta_j$.
(By using fixed-length bytes, we avoid the need to keep track of the bit offset in the buffer for 
variable length encodings.)
Figure \ref{fig:compression_example} illustrates the algorithm.

\subsection{Empirical Results}
\label{sec:empirical-results}
We give results on real and synthetic datasets.  
We use real data to show that our approach works in practice. We use synthetic data to demonstrate that the method scales effectively.

For the synthetic datasets, we generate an independent pseudorandom stream in which each item occurs only once.
We achieve this via the Fibonacci hashing scheme \cite[Section 6.4]{knuth1973art} and generate a 
unique stream for every trial.
We generate streams of cardinality $n$ ranging from $1$ to approximately $10^7$ .
For all comparisons to generic compressors, we use the $\texttt{zlib}$ \cite{zlib} and 
$\texttt{bzip2}$ \cite{bzip2} compressors available in Python.

\textbf{Experimental Setup.}
We use the default configuration of the Apache DataSketches Theta Sketch with $\texttt{lg\_k} = 12$
to initialize a sketch for each trial. 
Then we apply the sketch to the data stream for the trial. 
We execute $\texttt{trim}$ on the Theta sketch to produce a \kmv{} sketch with exactly
the $k=2^{12}$ smallest hash values from the stream.
For each input cardinality $n$, we run $256$ trials and plot the mean of the results.
To confirm that our implementation closely matches the theory, we plot the serialized sketch size 
in bytes against the input cardinality.
Also, to verify that our implementation is practical, we plot the time required to encode the sketch and write it to 
bytes.

\subsection{Comparison to entropy}
\label{sec:size-to-entropy}
Figure \ref{fig:implementation} shows the empirical results.
We plot the entropy estimate given by expression \eqref{eq:main_result}, converted to bytes by dividing by $8 \ln 2$, and the mean size of the compressed sketch.
For compression, we are interested in cases with $k < n$, so that the 
sketch keeps fewer items than are present in the stream.
We refer to that regime as \emph{estimation mode}.
The compressed sketch size is consistently close to the entropy and generally shrinks as $n$ grows.
The mean sketch size is only $4.760\%$ larger than the entropy.
Compressed sizes also concentrate extremely tightly: $99.7\%$ of the reported mean compressions 
are between $4.010\%$ and $5.774\%$ larger than the entropy, meanwhile \emph{all} reported mean compression sizes 
lie between $4.007\%$ and $5.775\%$ larger than the entropy.
This indicates that the average compression achieved by our implementation does not strongly deviate from the 
bound as the cardinality increases.

\input{compression.tex}

\subsection{Space Reduction}
Figure \ref{fig:implementation} illustrates that we obtain smaller sketches than either of the generic compressors;
thus achieving the practical aim of this work.
In order to understand the behavior of this reduction, we plot the the size ratio of compressed sketches to full, uncompressed sketches when in estimation mode.
Theorem \ref{hkn_thm} (proved later) predicts a saving of about $O(\log n)$ bits of space per key:
numerically, we observe that compressed sketches 
are about $85\%$ of the length of uncompressed sketches when $n$ only just exceeds $k$ 
(around the grey dashed line (\ref{estimation}) in Figure \ref{fig:implementation}).
As $n$ grows, the compressed sketches can be as small 
as $68\%$ of the size of uncompressed sketches.

The space savings increase with the logarithm of the input cardinality and can be roughly characterized as:
\begin{align}
    \textsf{Compression Ratio} &= \frac{\text{Compressed sketch size}}{\text{Uncompressed sketch size}} \label{eq:comp-defn} \\
    &\approx \frac{1.05}{8 \ln 2} \left[\ln(s) - \ln(n+1) - 1\right]
    \label{eq:empirical-space-saving}
\end{align}
which is easily derived by converting \eqref{eq:main_result} to bytes\footnote{Simply divide by $8 \ln(2)$.} and 
recalling from Section \ref{sec:size-to-entropy} that the sketch sizes in practice are roughly 
$5\%$ larger than the entropy expression \eqref{eq:main_result}, hence the factor $1.05$.
Then the uncompressed size of a \kmv{} sketch in bytes is roughly $64k / 8 = 8k$, since $64$-bit hash codes 
are stored.
The Theta sketch implementation reserves a further $24$ bytes for a header 
-- a small byte string that defines the sketch configuration parameters. So the overall space 
is $8\times2^{12} + 24 = 32792$ bytes.
The $\textsf{Compression Ratio}$ function is plotted in Figure \ref{fig:space-saving}, verifying that we observe an $O(\log n)$ space
per key reduction in the size when we compress the sketch.

\input{space_saving.tex}
\begin{table*}[htbp]
    \caption{Kasandr dataset results.  
    Sizes are in bytes and the ``Compression'' column takes the ratio of sizes between any of the compression methods and the standard method:
    $\textsf{CompressedSize}/\textsf{UncompressedSize}$.  
    Lower is better and the best compression ratio for each column is in bold.  
    We use $\texttt{lg\_k} = 12$ for all columns even though some columns have small cardinality.
    Good performance across all cardinalities is crucial because the cardinality of a column is not known in advance
    of data observation.
    Our method is the best in all columns; it always returning a smaller sketch unlike generic compressors.}
    \centering
    \begin{tabular}{@{}lllllllllll@{}}
        \toprule
                      & \multicolumn{2}{l}{Total}       & \multicolumn{2}{l}{offerid: $n=2158859$} & \multicolumn{2}{l}{userid: $n = 291485$} & \multicolumn{2}{l}{Merchant: $n=703$} & \multicolumn{2}{l}{category: $n=271$} \\ \midrule
                      & Size           & Compression    & Size                & Compression        & Size                & Compression        & Size              & Compression       & Size              & Compression       \\ \cmidrule(l){2-11} 
        Standard      & 73408          &                & 32792               &                    & 32792               &                    & 5640              &                   & 2184              &                   \\
        bzip2         & 73377          & 1.000          & 31697               & 0.967              & 32943               & 1.005              & 6145              & 1.090             & 2592              & 1.187             \\
        zlib          & 68937          & 0.939          & 30070               & 0.917              & 31021               & 0.946              & 5651              & 1.002             & 2195              & 1.005             \\ \midrule
        \textbf{Ours} & \textbf{54646} & \textbf{0.744} & \textbf{23058}      & \textbf{0.703}     & \textbf{24594}      & \textbf{0.750}     & \textbf{5019}     & \textbf{0.890}    & \textbf{1975}     & \textbf{0.904}    \\ \bottomrule
        \end{tabular}
    \label{tab:kasandr}
\end{table*}


\subsection{(De)Serialization Time}
Serialization is the process of writing the sketch to bytes. 
The binary file that stores sketches is a small, compact byte-string that can be efficiently written 
or transported out of memory to other machines.
Deserialization is the opposite process of reading a sketch from bytes and storing it as an object in memory. 
Large-scale systems are often based on multiple distributed compute nodes to exploit parallel computation.
Reporting systems often use thousands of sketches over small time periods that are aggregated into 
``merged'' sketches to summarize data over longer time periods.
Hence, it is vitally important that compressed sketches can be (de)serialized quickly. 
This experiment verifies that our encoding indeed satisfies this requirement.
We expect serialization \emph{with compression} to take at least as much time as without compression since 
we need to iterate through the sketch twice rather than once, but the difference is extremely small in 
\emph{absolute terms}.

Figure \ref{fig:serialization-time} illustrates that when in estimation mode, the compressed sketches 
only take a small amount more time to serialize than uncompressed sketches.
In absolute terms, it takes about $1.5$ microseconds to serialize \textbf{uncompressed} sketches but between 
$10$-$11$ microseconds to serialize \textbf{compressed} sketches.
Although this is just about a factor $7\times$ \emph{relative} slowdown, in absolute terms 
it is a negligible difference and will not inhibit the practicality of using compressed sketches.
The reason for this is that, from a systems perspective, the main bottlenecks are the shuffle operation 
in the distributed computation, or the file I/O.
Hence, it is a sensible tradeoff in large systems to spend more CPU time in generating small objects for 
communication in order to reduce potential I/O.
Our findings for deserialization time are much the same so are omitted.

\input{serialization_time.tex}

\textbf{Generic compressors.}  The generic compressors $\texttt{zlib}$ \cite{zlib} and 
$\texttt{bzip2}$ \cite{bzip2} \emph{always} take on the order of between $10^{-4}$ and $10^{-3}$ seconds to compress a sketch.
This is in stark contrast to Figure \ref{fig:serialization-time} where the compressed sketches take on the order of (at most) $10^{-5}$ seconds to compress
so the generic compressors are at least one order of magnitude slower than our method.

\textbf{Early Stopping.}
Another benefit of our approach is that a compressed sketch can be read sequentially from memory and merged into a sketch held in memory without 
deserializing the bytes in full.
This is because the encoding records the minimum number of leading zeros and reads the least hash keys first,
followed by its difference from the second key, $\delta_j$.
Hence, any subsequent keys are exactly reconstructed in sequence.
Merge operations benefit from early stopping in the compressed setting because subsequent values 
can be ignored once it is clear that any subsequently read hash key would exceed the 
$k$ smallest hash keys in the union of the sketches' hash keys.
This means that merging compressed sketches for arbitrary set operations can be performed more 
efficientlly than traversing the entire sketch.

\subsection{Real Data}
\begin{figure*}
    \caption{Comparison of all methods on the Kasandr dataset.  Chart title is the column name and cardinality.
    Left column is absolute size and the right hand column is the compression rate.
    In all cases our method is smaller than competitors.}
\begin{tikzpicture}

    \definecolor{darkorange25512714}{RGB}{255,127,14}
    \definecolor{darkgray176}{RGB}{176,176,176}
    \definecolor{lightgray204}{RGB}{204,204,204}
    \definecolor{steelcyan31119180}{RGB}{31,119,180}
    \definecolor{forestgreen4416044}{RGB}{44,160,44}
    
    \begin{groupplot}[group style={group size=2 by 4, vertical sep=1.125cm},  width=\textwidth/2, height=\textwidth/5,]
    \nextgroupplot[
    legend style={fill opacity=0.8, draw opacity=1, text opacity=1, draw=lightgray204},
    tick align=outside,
    tick pos=left,
    x grid style={darkgray176},
    xmajorgrids,
    xmin=-0.59, xmax=3.59,
    xtick style={color=black},
    xtick={0,1,2,3},
    xticklabels={Uncompressed,bzip2,zlib,Our method},
    y grid style={darkgray176},
    ylabel={Size (bytes)},
    ymajorgrids,
    ymin=0, ymax=34431.6,
    ytick style={color=black},
    title={\textbf{offerid}: $n=2158859$}
    ]
    \draw[draw=none,fill=darkorange25512714] (axis cs:-0.4,0) rectangle (axis cs:0.4,32792);
    \draw[draw=none,fill=black] (axis cs:0.6,0) rectangle (axis cs:1.4,31697);
    \draw[draw=none,fill=blue] (axis cs:1.6,0) rectangle (axis cs:2.4,30070);
    \draw[draw=none,fill=cyan] (axis cs:2.6,0) rectangle (axis cs:3.4,23058);
    
    \nextgroupplot[
    legend style={fill opacity=0.8, draw opacity=1, text opacity=1, draw=lightgray204},
    tick align=outside,
    tick pos=left,
    x grid style={darkgray176},
    xmajorgrids,
    xmin=-0.59, xmax=3.59,
    xtick style={color=black},
    xtick={0,1,2,3},
    xticklabels={Uncompressed,bzip2,zlib,Our method},
    y grid style={darkgray176},
    ymajorgrids,
    ymin=0, ymax=1.05,
    ytick style={color=black},
    ytick = {0, 0.25,0.5, 0.75,1.},
    title={\textbf{offerid}: $n=2158859$}
    ]
    \draw[draw=none,fill=darkorange25512714] (axis cs:-0.4,0) rectangle (axis cs:0.4,1);
    \draw[draw=none,fill=black] (axis cs:0.6,0) rectangle (axis cs:1.4,0.966607709197365);
    \draw[draw=none,fill=blue] (axis cs:1.6,0) rectangle (axis cs:2.4,0.916991949255916);
    \draw[draw=none,fill=cyan] (axis cs:2.6,0) rectangle (axis cs:3.4,0.703159307148085);

    \nextgroupplot[
    legend style={fill opacity=0.8, draw opacity=1, text opacity=1, draw=lightgray204},
    tick align=outside,
    tick pos=left,
    x grid style={darkgray176},
    xmajorgrids,
    xmin=-0.59, xmax=3.59,
    xtick style={color=black},
    xtick={0,1,2,3},
    xticklabels={Uncompressed,bzip2,zlib,Our method},
    y grid style={darkgray176},
    ylabel={Size (bytes)},
    ymajorgrids,
    ymin=0, ymax=34590.15,
    ytick style={color=black},
    title={\textbf{userid}: $n = 291485$}
    ]
    \draw[draw=none,fill=darkorange25512714] (axis cs:-0.4,0) rectangle (axis cs:0.4,32792);
    \draw[draw=none,fill=black] (axis cs:0.6,0) rectangle (axis cs:1.4,32943);
    \draw[draw=none,fill=blue] (axis cs:1.6,0) rectangle (axis cs:2.4,31021);
    \draw[draw=none,fill=cyan] (axis cs:2.6,0) rectangle (axis cs:3.4,24594);
    
    \nextgroupplot[
    legend style={fill opacity=0.8, draw opacity=1, text opacity=1, draw=lightgray204},
    tick align=outside,
    tick pos=left,
    x grid style={darkgray176},
    xmajorgrids,
    xmin=-0.59, xmax=3.59,
    xtick style={color=black},
    xtick={0,1,2,3},
    xticklabels={Uncompressed,bzip2,zlib,Our method},
    y grid style={darkgray176},
    ymajorgrids,
    ymin=0, ymax=1.05483502073677,
    ytick style={color=black},
    ytick = {0, 0.25,0.5, 0.75,1.},
    title={\textbf{userid}: $n = 291485$}
    ]
    \draw[draw=none,fill=darkorange25512714] (axis cs:-0.4,0) rectangle (axis cs:0.4,1);
    \draw[draw=none,fill=black] (axis cs:0.6,0) rectangle (axis cs:1.4,1.00460478165406);
    \draw[draw=none,fill=blue] (axis cs:1.6,0) rectangle (axis cs:2.4,0.945992925103684);
    \draw[draw=none,fill=cyan] (axis cs:2.6,0) rectangle (axis cs:3.4,0.75);
    
    \nextgroupplot[
    legend style={fill opacity=0.8, draw opacity=1, text opacity=1, draw=lightgray204},
    tick align=outside,
    tick pos=left,
    x grid style={darkgray176},
    xmajorgrids,
    xmin=-0.59, xmax=3.59,
    xtick style={color=black},
    xtick={0,1,2,3},
    xticklabels={Uncompressed,bzip2,zlib,Our method},
    y grid style={darkgray176},
    ylabel={Size (bytes)},
    ymajorgrids,
    ymin=0, ymax=6452.25,
    ytick style={color=black},
    title={\textbf{Merchant}: $n=703$}
    ]
    \draw[draw=none,fill=darkorange25512714] (axis cs:-0.4,0) rectangle (axis cs:0.4,5640);
    \draw[draw=none,fill=black] (axis cs:0.6,0) rectangle (axis cs:1.4,6145);
    \draw[draw=none,fill=blue] (axis cs:1.6,0) rectangle (axis cs:2.4,5651);
    \draw[draw=none,fill=cyan] (axis cs:2.6,0) rectangle (axis cs:3.4,5019);
    
    \nextgroupplot[
    legend style={fill opacity=0.8, draw opacity=1, text opacity=1, draw=lightgray204},
    tick align=outside,
    tick pos=left,
    x grid style={darkgray176},
    xmajorgrids,
    xmin=-0.59, xmax=3.59,
    xtick style={color=black},
    xtick={0,1,2,3},
    xticklabels={Uncompressed,bzip2,zlib,Our method},
    y grid style={darkgray176},
    ymajorgrids,
    ymin=0, ymax=1.14401595744681,
    ytick style={color=black},
    ytick = {0, 0.25,0.5, 0.75,1.},
    title={\textbf{Merchant}: $n=703$}
    ]
    \draw[draw=none,fill=darkorange25512714] (axis cs:-0.4,0) rectangle (axis cs:0.4,1);
    \draw[draw=none,fill=black] (axis cs:0.6,0) rectangle (axis cs:1.4,1.0895390070922);
    \draw[draw=none,fill=blue] (axis cs:1.6,0) rectangle (axis cs:2.4,1.00195035460993);
    \draw[draw=none,fill=cyan] (axis cs:2.6,0) rectangle (axis cs:3.4,0.889893617021277);
    
    \nextgroupplot[
    legend style={fill opacity=0.8, draw opacity=1, text opacity=1, draw=lightgray204},
    tick align=outside,
    tick pos=left,
    x grid style={darkgray176},
    xmajorgrids,
    xmin=-0.59, xmax=3.59,
    xtick style={color=black},
    xtick={0,1,2,3},
    xticklabels={Uncompressed,bzip2,zlib,Our method},
    y grid style={darkgray176},
    ylabel={Size (bytes)},
    ymajorgrids,
    ymin=0, ymax=2721.6,
    ytick style={color=black},
    title={\textbf{category}: $n=271$}
    ]
    \draw[draw=none,fill=darkorange25512714] (axis cs:-0.4,0) rectangle (axis cs:0.4,2184);
    \draw[draw=none,fill=black] (axis cs:0.6,0) rectangle (axis cs:1.4,2592);
    \draw[draw=none,fill=blue] (axis cs:1.6,0) rectangle (axis cs:2.4,2195);
    \draw[draw=none,fill=cyan] (axis cs:2.6,0) rectangle (axis cs:3.4,1975);
    
    \nextgroupplot[
    legend style={fill opacity=0.8, draw opacity=1, text opacity=1, draw=lightgray204},
    tick align=outside,
    tick pos=left,
    x grid style={darkgray176},
    xmajorgrids,
    xmin=-0.59, xmax=3.59,
    xtick style={color=black},
    xtick={0,1,2,3},
    xticklabels={Uncompressed,bzip2,zlib,Our method},
    y grid style={darkgray176},
    ymajorgrids,
    ymin=0, ymax=1.24615384615385,
    ytick style={color=black},
    ytick = {0, 0.25,0.5, 0.75,1.},
    title={\textbf{category}: $n=271$}
    ]
    \draw[draw=none,fill=darkorange25512714] (axis cs:-0.4,0) rectangle (axis cs:0.4,1);
    \draw[draw=none,fill=black] (axis cs:0.6,0) rectangle (axis cs:1.4,1.18681318681319);
    \draw[draw=none,fill=blue] (axis cs:1.6,0) rectangle (axis cs:2.4,1.00503663003663);
    \draw[draw=none,fill=cyan] (axis cs:2.6,0) rectangle (axis cs:3.4,0.904304029304029);
    \end{groupplot}
    
    \end{tikzpicture}
    \label{fig:kasandr-bar-charts}
\end{figure*}
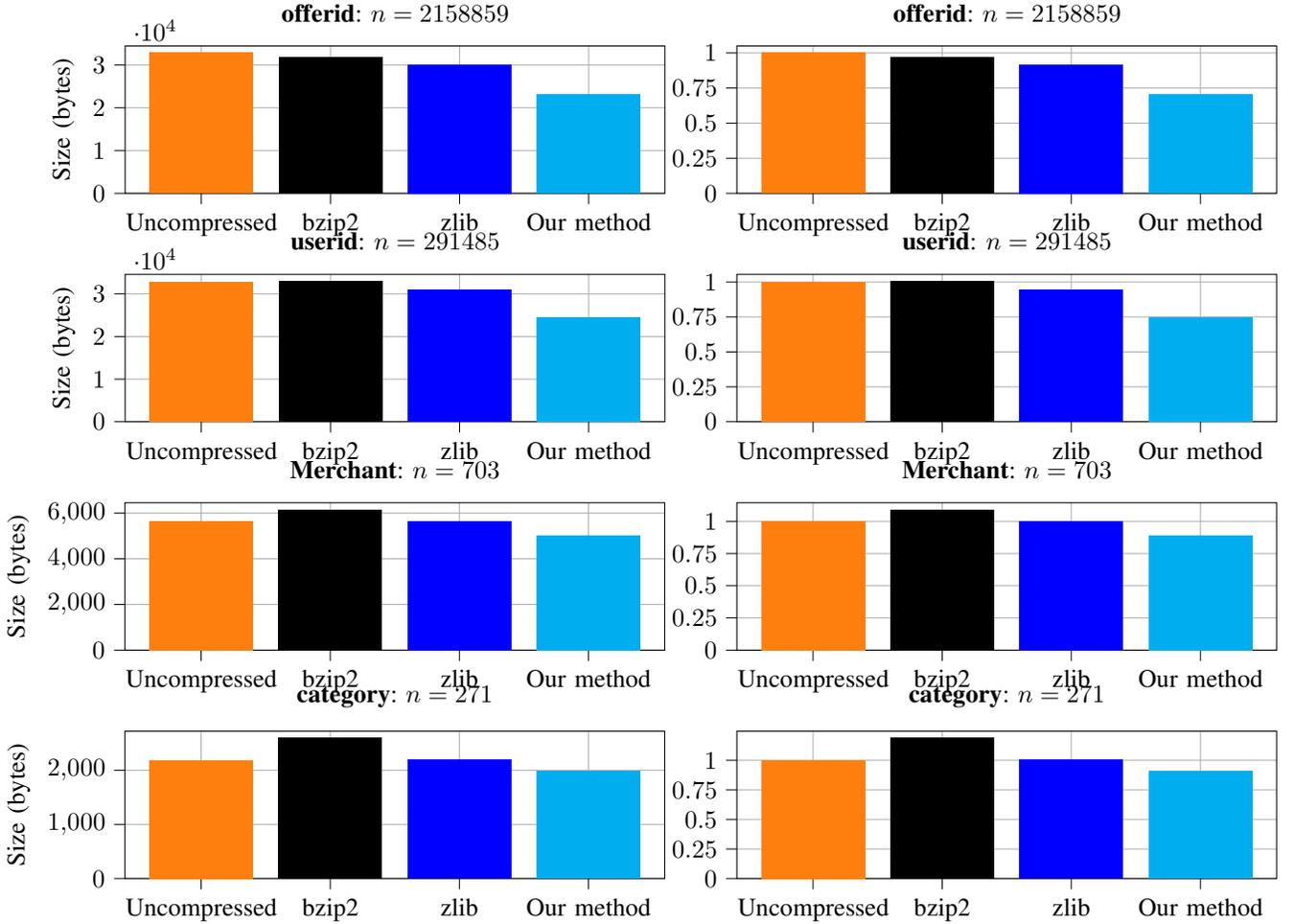
We test the compression scheme on the Kasandr dataset \cite{sidana2017kasandr,kasandr385} which is a 
dataset of user interactions with online advertisements.
The dataset has about $17.76$ million records in total and has four columns with 
cardinalities varying from the hundreds to the millions as described in the header of Table \ref{tab:kasandr}.
The total number of records exceeds the cardinality in any column which means that some values in 
the column repeat.  
However, this does not affect any of the compression schemes because they take as input a \kmv{} sketch
which is invariant to frequency of items in the stream.
Thus each column can be viewed as a single vertical slice of Figure \ref{fig:implementation} with $n$ being the column 
cardinality.

We use the same \kmv{} setup as described in Section \ref{sec:empirical-results} and measure the size of the uncompressed sketch 
and the compressed sketches with compression performed using zlib, bzip2, and our method.
Figure \ref{fig:kasandr-bar-charts} shows the results.
Again, we timed the compression techniques over thousands of trials and found that the generic compressors took much more time:
on high cardinality columns, the generic compressors took on the order of $10^{-4}-10^{-3}$ seconds compared to 
on the order of $10^{-6}$ seconds for our method, which is only marginally more than serialization without compression.
For small cardinality columns an order of magnitude less time is needed for all methods but the relationships remain the 
same: using our compression method is competitive with serialization without compression and is much faster than the generic compressors.\footnote{
    These timings were performed using $\texttt{\%\% timeit}$ in Python. 
    There is essentially no difference in the timings between the Python and C++ bindings for the Apache DataSketches methods.
}

Our findings on the Kasandr dataset are consistent with the synthetic data.
In the left column of Figure \ref{fig:kasandr-bar-charts} we plot the absolute size of the sketches in bytes and in the right column we plot the
compression ratio (as in Equation \eqref{eq:comp-defn}) for each sketch.
A lower compression ratio is better and the uncompressed sketch trivially has a compression ratio of $1$.
The plots are ordered with the highest cardinality column at the top and the smallest at the bottom. 

We find that on all columns, our compression scheme offers the best savings.
Indeed, our method always reduces the size of the input whereas the generic compressors may inflate the sketch 
at small cardinalities.
Our compression scheme is best when the cardinality is largest and consistently outperforms the other compression schemes.
The measurements from which this plot is derived are given in Table \ref{tab:kasandr}.

\section{Successive Differences -- Entropy Estimate} \label{hbs_sec}
The next few sections focus on entropy and entropy estimation for the distribution over which we compress sketches -- the $k$ minimum values from $n$ drawn uniformly at random from the space of $s$ hash codes.

\subsection{Warm-up}
As a warm-up (and for comparison), consider the entropy for a \emph{uniform distribution} over space of hash codes. 
The definition of entropy is $H = - \sum_i p_i \ln p_i,$
where the sum is over all possible outcomes $i$.
For a uniform distribution over $s$ outcomes, the entropy is
\begin{equation}
H_{U} = - \sum_{i = 0}^{s - 1} \frac{1}{s} \ln \frac{1}{s} = - \ln \frac{1}{s} \left(\sum_{i = 0}^{s - 1} \frac{1}{s}\right) = - \ln \frac{1}{s}  = \ln s. \label{hu}
\end{equation}

Next, consider the entropy of the distribution of the smallest value and successive gaps for $n$ hash codes drawn uniformly at random from a space of $s$ hash codes: $0$ to $s - 1$. In essence, we want to analyze how much entropy is reduced by ordering the size-$n$ sample. To simplify the analysis, we treat the sampling process as drawing each value i.i.d. from $U[0,1]$, then mapping any value in $\left[\frac{i}{s}, \frac{i+1}{s}\right)$ to hash code $i$. For $n$ order statistics drawn uniformly at random from $[0, 1]$, the distribution of the minimum value and of each successive gap is $\Bd(1, n)$ \cite{david03}.

\begin{thm} \label{hbs_thm}
Let $H_{B|S}$ be the entropy of the process: draw a value from distribution $\Bd(1, n)$ then map the value to the maximum integer $i \in [0, s - 1]$ such $\frac{i}{s}$ is at most the value. Then
\begin{equation}
H_{B|S} \approx \ln s - \ln n + \frac{n - 1}{n}. \label{hbs}
\end{equation}
\end{thm}

(See Appendix \ref{sec:deferred-proofs} for the proof.) Comparing $H_{B|S}$ from Theorem \ref{hbs_thm} to $H_{U}$ from Expression \ref{hu}, we see that the difference is about $\ln n$. Intuitively, this is the reduction in entropy from ordering the samples -- the information needed to communicate or store the hash codes is reduced by about $\ln n$ per sample since we do not require them to be communicated or stored in their original order. We could restore the ordering by sending or storing $O(\ln n)$ bits for each sample -- encoding a value from 0 to $n - 1$ that indicates which draw produced each hash code. 

Theorem \ref{hbs_thm} indicates that it is possible to compress initial values or successive differences enough to reduce their storage by about $\log_2 n$ bits from the $\log_2 s$ bits for the direct encoding of hash codes. For example, if we set $s=2^{64}$ and input cardinality is on the order of billions, $n \approx 2^{30}$, then the entropy is approximately $34$ bits -- a reduction by almost one half. Note that the results are per initial value or successive difference, so we can hope to save about $k \ln n$ bits per sketch if each sketch communicates the $k$ least hash codes for the values in a data source.

From another point of view, the number of distinct values $n$ in the data source increases, the average of the least of their hash codes and the average of successive differences between their ordered hash codes shrinks as 
$O(\frac{1}{n})$. (For our $\Bd(1, n)$ distribution, the mean is $\frac{1}{n + 1}$, and the variances decrease as well, making those values more predictable in an information-theoretic sense.) So initial values and successive differences are more compressible as the number of distinct values increases. 

Next, we note that the approximation error in Theorem \ref{hbs_thm} is small:

\begin{thm} 
    \label{hbs_err_thm}
    Under the assumptions of Theorem \ref{hbs_thm}, the approximation 
    \begin{equation*}
        H_{B|S} \approx \ln s - \ln n + \frac{n - 1}{n}
    \end{equation*}
    incurs error $O(\frac{n \ln n}{s})$.
\end{thm}
(See Appendix \ref{sec:deferred-proofs} for the proof.)

\section{Leading Zeros in Successive Differences} \label{leading_zero_sec}
Initial values and successive differences between hash codes have the form $i \in [0, s-1]$ -- an unsigned integer. One method to compress those values is to determine the minimum number of leading zeros for the bitwise representation of the $i$ values, indicate that value at the start of the message, then remove that many leading zeros from each value and transmit the remaining bits (as illustrated in Figure \ref{fig:compression_example}). 

To get a sense of the number of leading zeros, consider that the average $i$ value for the initial value and successive differences is about $\frac{1}{n}$ of $s$. Of the $\log_2 s$ bits for integers $0$ to $s - 1$, the value $\frac{s}{n}$ has about $\log_2 n$ leading zeros. But we are concerned with the minimum number of leading zeros over the $k$ values.

To get a sense of that value, consider that for uniform random values in $[0,1]$ the probability of an initial value greater than $\frac{c}{n - 1}$ is \cite{baxbax22}:
\begin{equation*}
\left(1 - \frac{c}{n - 1}\right)^n \approx e^{-c}.
\end{equation*}
So the probability of an initial value greater than $\frac{2 \ln n}{n}$ is about $O(\frac{1}{n^2})$. 

For successive differences, imagine partitioning $[0,1]$ into about $\frac{n}{c}$ successive regions of size-$\frac{c}{n}$ each. For a successive difference to be at least $\frac{2c}{n}$, the gap between the hash codes must be a superset of one of the size-$\frac{c}{n}$ regions. So an upper bound on the probability that any one such region has none of the $n$ hash codes is an upper bound on the probability of a successive difference at least $\frac{2c}{n}$. 

Following the logic from \cite{baxbax22}, the probability that a specific region of length $\frac{c}{n}$ is empty is 
\begin{equation*}
\left(1 - \frac{c}{n}\right)^n \approx e^{-c}.
\end{equation*}
By the sum bound on the union of probabilities, the probability that any of the approximately $\frac{n}{c}$ regions are empty is
\begin{equation*}
\approx \frac{n}{c} \left(1 - \frac{c}{n}\right)^n \approx \frac{n}{c e^{c}}.
\end{equation*}
For $c = 2 \ln n$, we get a probability of any empty region that is about $O(\frac{1}{n^2})$. 
So the probability that no successive difference is $\frac{2c}{n} = \frac{4 \ln n}{n}$ is 
about $\frac{1}{2 n \ln n}$.

So, for $n$ over a thousand, the probability of an initial value or successive difference greater than $\frac{4 \ln n}{n}$ is under about $\frac{1}{14\,000}$. With high probability, then, the values we transmit will all be less than $\frac{4 s \ln n}{n}$, giving us about $\log_2 n - \log_2 \ln n - 2$ leading zero bits. For $n$ one million, that is about 14 bits of reduced length per value in the sketch.

\section{Entropy for Subsets of Least Elements} \label{hkn_sec}
The entropy results for individual successive differences from Section \ref{hbs_sec} motivate compression by removing leading zeros, as discussed in Section \ref{leading_zero_sec}. 
But how well do they justify such an approach? 
Even though removing leading zeros comes close to achieving the compression limit based on entropy for individual successive differences, there remains the possibility that the entropy for the least $k$ of size-$n$ subset of a size-$s$ set has substantially less entropy than the sum of entropies for successive differences, indicating that it is possible to achieve much better compression by some other means. 
In this section, we show that is untrue. 
Instead, we show that the entropy of the least $k$ elements from a size-$n$ subset is close to the sum of entropies for $k$ individual successive differences. 

Assuming hash codes are selected uniformly at random without collisions for $n$ distinct values 
and the least $k$ of the size-$n$ set of hash codes is to be communicated in a data sketch, 
we estimate a lower bound on the number of bits needed to encode the size-$k$ subset of hash codes. 
To do that, we will appeal to Shannon's source coding theorem, so we need to estimate the entropy for the distribution 
over size-$k$ subsets of a size-$s$ set if the size-$k$ subset is drawn by first drawing a size-$n$ subset at random, 
then taking the least $k$ elements from the size-$n$ subset. 
To be concrete, let the size-$s$ set be the integers 1 to $s$, but the result applies to any set 
with a method to order elements, so that we can select the least $k$ of $n$ elements. 

\begin{thm} \label{hkn_thm}
Let $H_{K|N}$ be the entropy of the distribution of the subset of $k$ least elements of a size-$n$
subset selected uniformly at random without replacement from a size-$s$ set, with $n \ll s$. Then
\be
H_{K|N} \approx k [\ln s - \ln (n + 1) +1]. \label{hkn}
\ee
\end{thm}

This indicates that about $\log_2 s - \log_2 (n + 1) + \log_2(e)$ bits are required per hash code, on average, in a sketch that communicates the $k$ least of $n$ hash codes. 
That is a saving of $\log_2 (n + 1) - \log_2(e)$ bits per hash code compared to transmitting the hash codes without compression, since then each hash code would require $\log_2 s$ bits.

\begin{proof}
    Let $p_j$ be the probability that a random size-$n$ subset of a size-$s$ set has $k$th least element $j$. Then
    \begin{equation}
    p_j = \frac{{{j - 1}\choose{k - 1}}{{1}\choose{1}}{{s - j}\choose{n - k}}}{{{s}\choose{n}}}, \label{pj_def}
    \end{equation}
    since, among the $n$ elements chosen from $s$ elements, $k - 1$ must come from the first $j - 1$ elements, element $j$ must be selected, and the other $n - k$ must be selected from the last $s - j$ elements. 
    
    Given that $j$ is the $k$th least element, each size-$k$ subset of elements has probability:
    \begin{equation}
    p_k = \frac{1}{{{j - 1}\choose{k - 1}}}, \label{pk_def}
    \end{equation}
    since the number of such size-$k$ subsets is the number of ways to select $k - 1$ elements from the $j - 1$ elements of the the size-$s$ set that are less than $j$, and each such size-$k$ subset is equally likely. 
    
    The entropy is
    \begin{equation*}
    H_{K|N} = - \sum_{j = k}^{s} {{j - 1}\choose{k - 1}} p_j p_k \ln (p_j p_k),
    \end{equation*}
    since there are ${{j - 1}\choose{k - 1}}$ size-$k$ subsets with greatest element $j$, and each has probability $p_j p_k$ -- the probability that the greatest element is $j$, times the probability of the set given that the greatest element is $j$. (The sum begins with $j = k$ rather than $j = 0$ because the $k$th least sample must be at least the $k$th least element, so $p_j = 0$ for $j < k$.) Since ${{j - 1}\choose{k - 1}}$ is the inverse of $p_k$:
    \begin{equation*}
    H_{K|N} = - \sum_{j = k}^{s} p_j \ln (p_j p_k). 
    \label{hkn_one_term}
    \end{equation*}
    Separating terms in the logarithm,
    \be
    H_{K|N} = - \sum_{j = k}^{s} p_j \ln p_j - \sum_{j = k}^{s} p_j \ln p_k. \label{two_terms}
    \ee
    We now separate the analysis and deal with the first sum and second sum separately.
    
    \textbf{First sum \eqref{two_terms}.} This is the entropy for the distribution of $j$ -- the $k$th element of a size-$n$ subset of the integers in $[1,s]$. Using the reasoning from the proof of Theorem \ref{hbs_thm}, that entropy is approximately $\ln s$ plus the differential entropy of the distribution of the $k$th least among $n$ i.i.d. samples from a uniform distribution over $[0,1]$. Since that has a $\Bd(k, n - k + 1)$ distribution (see \cite{david03}), 
    it has differential entropy (\cite{lazo78,wiki_beta}) $H_{B_{k}}$:
    
    \begin{equation}
        \begin{split}
            H_{B_{k}} = &\ln \mathrm{B}(k, n - k + 1) - (k - 1) \psi(k)  \\
            &- (n - k) \psi(n - k + 1) + (n - 1) \psi(n + 1).
        \end{split}
    \label{eq:big-diff-entropy}
    \end{equation}
    Using equalities $\mathrm{B}(\alpha, \beta) = \frac{\Gamma(\alpha)\Gamma(\beta)}{\Gamma(\alpha + \beta)}$ and $\Gamma(\alpha) = (\alpha - 1)!$
    \be
    \ln \mathrm{B}(k, n - k + 1) = \ln \frac{(k - 1)! (n - k)!}{n!}. \label{from_b}
    \ee
    Using Stirling's approximation: $\ln n! \approx n \ln n - n$, then dropping a term of $-1$:
    \be
    \eqref{from_b} \approx - n \ln n + (n - k) \ln (n - k) + (k - 1) \ln (k - 1). \label{to_b}
    \ee
    For the remaining terms in \eqref{eq:big-diff-entropy} we apply the approximation
    $\psi(z) \approx \ln z$:
    \be
    H_{B_{k}}\approx - (k - 1) \ln k - (n - k) \ln (n - k + 1) + (n - 1) \ln (n + 1).
    \label{eq:remaining-terms}
    \ee
    Combining \eqref{to_b} and \eqref{eq:remaining-terms}, 
    the differential entropy $H_{B_{k}}$ of $\Bd(k, n - k + 1)$ is
    
    \begin{equation}
        \begin{split}
            H_{B_{k}} &\approx (n - 1) \ln (n + 1) - n \ln n + (n - k) \ln (n - k)  \\ 
            &- (n - k) \ln (n - k + 1) + (k - 1) \ln (k - 1) \\
            &- (k - 1) \ln k.
        \end{split}
        \label{eq:diff-entropy-approximation}
    \end{equation}
    Collecting terms we obtain
    \begin{equation}
        \begin{split}
            \eqref{eq:diff-entropy-approximation} 
            &= \ln \left(\frac{n + 1}{n}\right)^{n}  - \ln \left(\frac{k}{k - 1}\right)^{k - 1}  \\ 
            &-\ln \left(\frac{n - k + 1}{n - k}\right)^{n - k}- \ln (n + 1).
        \end{split}
        \label{eq:diff-entropy-log-format}
    \end{equation} 
    Now, using the approximation $\left(\frac{t + 1}{t}\right)^{t} \approx e$, 
    the first three terms of \eqref{eq:diff-entropy-log-format} are approximately $\ln e = 1$ and two of them cancel. 
    Hence, by dropping the remaining $-1$ additive term, 
    the differential entropy of $\Bd(k, n - k + 1)$ is
    \begin{equation*}
    H_{B_{k}} \approx - \ln (n + 1).
    \end{equation*}
    Recalling the $\ln s$ contribution, the first term in \eqref{two_terms} is
    \begin{equation}
    - \sum_{j = k}^{s} p_j \ln p_j \approx \ln s - \ln (n + 1).
    \label{eq:first_term_final}
    \end{equation}
    
    \textbf{Second sum \eqref{two_terms}.} We recognise that the term, $- \sum_{j = k}^{s} p_j \ln p_k$
    is the expectation over the distribution $j$
    -- the $k$th least sample of $n$ ordered samples from the integers $1$ to $s$ -- of:
    \begin{equation*}
    - \ln p_k = \ln {{j - 1}\choose{k-1}}.
    \end{equation*}
    Since $j - 1 \gg k - 1$ for the likely values of $j$, use the approximation:
    \begin{equation*}
    {{j - 1}\choose{k-1}} \approx \frac{(j - 1)^{k - 1}}{(k - 1)!}.
    \end{equation*}
    Then 
    \be
    - \sum_{j = k}^{s} p_j \ln p_k \approx (k - 1) \sum_{j = k}^{s} p_j \ln (j - 1) - \sum_{j = k}^{s} p_j \ln (k - 1)! .
    \label{eq:second-term-split-in-two}
    \ee
    For the first term of \eqref{eq:second-term-split-in-two}, we let $X \sim \Bd(k, n - k + 1)$. 
    Then the distribution of $X$ is very close to that of $\frac{j - 1}{s}$ for large $s$. 
    So
    \begin{align}
    (k - 1) \sum_{j = k}^{s} p_j \ln (j - 1) 
        &\approx (k - 1) \mean \ln s X  \nonumber \\
        &= (k - 1) \left[\ln s + \mean \ln X\right] \label{eq:intermediate_first_sum}
    \end{align}
    But for $X \sim \Bd(\alpha, \beta)$, $\mean \ln X = \psi(\alpha) - \psi(\alpha + \beta)$, so
    \be
    \eqref{eq:intermediate_first_sum} \approx (k - 1) \left[\ln s - \psi(n + 1) + \psi(k)\right]
    \ee
    Since $\psi(z) \approx \ln z$, we obtain
    \be
    \eqref{eq:intermediate_first_sum} \approx (k - 1) [\ln s - \ln (n + 1) + \ln k].
    \label{eq:second-sum-first-term-bound}
    \ee
    
    Now for second summation of \eqref{eq:second-term-split-in-two} first, we  note that 
    \begin{equation}
        \sum_{j = k}^{s} p_j \ln (k - 1)! = \ln (k - 1)! \sum_{j = k}^{s} p_j     
        \label{eq:final-eq-to-bound}
    \end{equation}
    
    \be
    \eqref{eq:final-eq-to-bound} = \ln (k - 1)! \approx (k - 1) \ln (k - 1) - (k - 1),
    \label{eq:second-sum-second-term-bound}
    \ee
    by Stirling's approximation and again noting that $\sum_{j=k}^s p_j = 1$ as $p_j = 0$ for $j < k$. 
    
    
    We combine \eqref{eq:second-sum-first-term-bound} and 
    \eqref{eq:second-sum-second-term-bound} to obtain
    \begin{align}
        \eqref{eq:second-term-split-in-two} &\approx
        (k - 1) [\ln s - \ln (n + 1) + \ln k] \nonumber  \\
        &- (k - 1) \ln (k - 1) + (k - 1)  \nonumber \\ 
        &\approx (k - 1) [\ln s - \ln (n + 1) + 1]. \label{eq:second_sum_final}
    \end{align}
    Finally, we can return to the quantity we sought to bound, \eqref{two_terms}; 
    using \eqref{eq:first_term_final} for the first sum and \eqref{eq:second_sum_final} for the second
    sum yields
    \begin{align*}
        H_{K|N} &\approx \ln s - \ln (n + 1) + (k - 1) [\ln s - \ln (n + 1) + 1] \\ 
        &\approx k [\ln s - \ln (n + 1) + 1].
    \end{align*}
\end{proof}

The approximation error in Theorem \ref{hkn_thm} is small:
\begin{thm} \label{hkn_err_thm}
    Under the assumptions of Theorem \ref{hkn_thm}, the approximation
    \begin{equation*}
        H_{K|N} \approx k [\ln s - \ln (n + 1) +1]
    \end{equation*}
    incurs error O($\max(\frac{n \ln n}{s}, \frac{\ln k}{k}$)) bits per hash code in the sketch.
\end{thm}
(See Appendix \ref{sec:deferred-proofs} for the proof.)

\section{Discussion} \label{discussion_sec}
For compressing the least $k$ of $n$ random keys from a size-$s$ set of keys, 
we have shown (Theorem \ref{hkn_thm}) that the entropy is approximately $\log_2 s - \log_2 (n + 1) + \log_2(e)$
bits per key, indicating that we cannot hope to compress those keys in a way that saves more 
than about $\log_2 (n + 1) + \log_2(e)$ bits per key, according to Shannon's source coding theorem. 
We have also shown (Theorem \ref{hbs_thm}) that the entropy of the distribution of the least key
and the distribution of differences between successive keys is about 
$\log_2 s - \log_2 n + \frac{n-1}{n} \log_2(e)$, indicating that compressing keys based on the initial
key value and differences between successive keys can provide nearly optimal compression. 
For one such method of compression, removing leading zeros from the initial key and successive differences, 
we have shown (Section \ref{leading_zero_sec}) that we can hope to save about 
$\log_2 n - \log_2 \ln n - 2$ bits per key, since that is a lower bound on the minimum number of leading zeros, with high probability.

We have presented a highly-performant implementation of a compression scheme based on removing leading zeros in the initial hash key and successive differences.
The implementation is faithful to the theoretical findings from Theorems \ref{hbs_thm} and \ref{hkn_thm} 
and compresses sketches to only a very small percentage larger than the entropy (scaled to bytes).
Though the compression requires a small amount of additional computation, it is extremely quick in absolute terms, yielding a practical compression scheme.

In the future, it would be interesting to explore other compression methods for the least $k$ of $n$ keys. 
For an extension of removing leading zeros, imagine writing each of the $k$ items -- 
the initial key and successive differences between keys -- as a bit string in a column, with the leading bit on top, 
and columns from left to right. Then encode the bit sequence that read across each row, starting from the top, 
so the sequence of the most significant bits, then the sequence of the second-most significant bits, and so on. 
If all items have leading zeros, then the leading rows will be all-zero rows, and we can simply transmit how 
many of the rows are all-zero rows, as in removing leading zeros. 
But the next few rows, while not being all zeros, may be zero-heavy, making them compressible. 
It would be interesting to explore whether this could lead to significant savings in storage with quick computation.


\newpage
\bibliographystyle{unsrt}
\bibliography{sample}
\newpage
\begin{appendices}

\section{Deferred Proofs}
\label{sec:deferred-proofs}

\paragraph*{Proof of Theorem \ref{hbs_thm}}
\begin{proof}
    Let $F$ be the cdf and let $f$ be the pdf for $\Bd(1, n)$. 
    Then the entropy is 
    \begin{align*}
        &H_{B|S} = \\
        &- \sum_{i = 0}^{s - 1}  \left[F\left(\frac{i + 1}{s}\right) - F\left(\frac{i}{s}\right)\right]
    \ln \left[F\left(\frac{i + 1}{s}\right) - F\left(\frac{i}{s}\right)\right],
    \end{align*}
    since values in $\left[\frac{i}{s}, \frac{i + 1}{s}\right)$ map to $\frac{i}{s}$ in discretization. 
    
    By the Fundamental Theorem of Calculus, we have $f(i/s) = F'(i/s)$ and using the estimate:
    \begin{equation*}
    F'\left(\frac{i}{s}\right) \approx \frac{F\left(\frac{i + 1}{s}\right) - F\left(\frac{i}{s}\right)}{\frac{1}{s}},
    \end{equation*}
    thus
    \begin{equation*}
        F\left(\frac{i + 1}{s}\right) - F\left(\frac{i}{s}\right) \approx \frac{1}{s} f\left(\frac{i}{s}\right).    
    \end{equation*}
    Hence,
    \begin{equation*}
    H_{B|S} \approx - \sum_{i = 0}^{s - 1} \frac{1}{s} f\left(\frac{i}{s}\right) 
    \ln \left[\frac{1}{s} f\left(\frac{i}{s}\right)\right].
    \end{equation*}
    Separating the terms in the logarithm,
    \be
    = \ln s \sum_{i = 0}^{s - 1} \frac{1}{s} f\left(\frac{i}{s}\right) - 
    \sum_{i = 0}^{s - 1} \frac{1}{s} f\left(\frac{i}{s}\right) \ln f\left(\frac{i}{s}\right).
    \label{eq:thm1-entropy-est}
    \ee
    Assume $s$ is large, set $x = \frac{i}{s}$ and $\mathrm{d}x = \frac{1}{s}$.
    Then the first term of \eqref{eq:thm1-entropy-est} is
    \begin{equation*}
    \sum_{i = 0}^{s - 1} \frac{1}{s} f\left(\frac{i}{s}\right) \approx \int_{x = 0}^1 f(x) \mathrm{d}x = 1,
    \end{equation*}
    and the second term of \eqref{eq:thm1-entropy-est} is 
    \begin{equation*}
    \sum_{i = 0}^{s - 1} \frac{1}{s} f\left(\frac{i}{s}\right) 
    \ln f\left(\frac{i}{s}\right) \approx \int_{x = 0}^1 f(x) \ln f(x) \mathrm{d}x.
    \end{equation*}
    Then
    \begin{equation}
    H_{B|S} \approx \ln s -  \int_{x = 0}^1 f(x) \ln f(x) \mathrm{d}x. \label{hbs_inter}
    \end{equation}
    The integral term on the right of \eqref{hbs_inter} is the differential entropy of 
    $\Bd(1, n)$ which we denote $H_{B_{1}}$:
    \begin{equation*}
    H_{B_{1}} = -  \int_{x = 0}^1 f(x) \ln f(x) \mathrm{d}x.
    \end{equation*}
    In general, the differential entropy of $\Bd(\alpha, \beta)$ is \cite{lazo78,wiki_beta}
    \begin{equation*}
    \ln \mathrm{B}(\alpha, \beta) - (\alpha - 1) \gamma(\alpha) - (\beta - 1)\gamma(\beta) + (\alpha + \beta - 2) \gamma(\alpha + \beta),
    \end{equation*}
    where $\mathrm{B}$ is the beta function and $\psi$ is the digamma function \cite{abramowitz72,spanier87,weisstein,wiki_digamma}. For us, $\alpha = 1$ and $\beta = n$, so
    \begin{equation*}
    H_{B_{1}} = \ln \mathrm{B}(1, n) - (n - 1)\psi(n) + (n - 1) \psi(n + 1).
    \end{equation*}
    Since the beta function is
    \begin{equation*}
    \mathrm{B}(\alpha, \beta) = \int_0^1 t^{\alpha - 1} (1 - t)^{\beta - 1} \mathrm{d}t,
    \end{equation*}
    we have
    \be
    H_{B_{1}} = \ln \int_0^1 (1 - t)^{n - 1} \mathrm{d}t + (n - 1) (\psi(n + 1) - \psi(n)).
    \label{eq:diff-entropy}
    \ee
    The integral in \eqref{eq:diff-entropy} is $\frac{1}{n}$. Note that $\psi(n + 1) - \psi(n) = \frac{1}{n}$.  So
    \begin{equation*}
    H_{B_{1}} = \ln \frac{1}{n} + \frac{n - 1}{n} = - \ln n + \frac{n - 1}{n}.
    \end{equation*}
    Substituting \eqref{eq:diff-entropy} into \eqref{hbs_inter}:
    \begin{equation*}
    H_{B|S} \approx \ln s - \ln n + \frac{n - 1}{n}.
    \end{equation*}
\end{proof}

\paragraph*{Proof of Theorem \ref{hbs_err_thm}}

\begin{proof}
In Theorem \ref{hbs_thm}, we use the approximation:
\begin{equation*}
F\left(\frac{i + 1}{s}\right) - F\left(\frac{i}{s}\right) \approx \frac{1}{s} f\left(\frac{i}{s}\right),
\end{equation*}
which is equivalent to the standard linear approximation (which is a Taylor series \cite{apostol91,stewart12} with a linear term):
\begin{equation*}
F(x + \Delta) = F(x) + \Delta F'(x) + \epsilon(x),
\end{equation*}
with $x = \frac{i}{s}$ and $\Delta = \frac{1}{s}$. The second derivative provides an error bound (based on the Lagrange remainder for a Taylor series \cite{apostol91,stewart12}):
\begin{align}
|\epsilon(x)| &\leq |U| \frac{\Delta^2}{2} = \frac{|U|}{2s^2}.
\end{align}
where $U$ bounds the second derivative ($F''$) over $[x, x + \Delta]$. 
The average of $F''$ over $[0,1]$ is its integral over $[0, 1]$, 
namely, $F'(1) - F'(0)$. 
Since $F' = f$ is the pdf of $\Bd(1, n)$, 
\begin{equation*}
F'(x) = \frac{\Gamma(n + 1)}{\Gamma(1) \Gamma(n)} \left(1 - x\right)^{n - 1}
\end{equation*}
with $\Gamma(z) = (z - 1)!$, so
\begin{equation*}
F'(x) = n \left(1 - x\right)^{n - 1}
\end{equation*}
and the average of $F''$ is 
\begin{equation*}
F'(1) - F'(0) = n - 0 = n.
\end{equation*}
So the average error per linear approximation is at most $\frac{n}{2s^2}$. 
Since we make $O(s)$ such approximations from the size of the $\Delta$ discretizations, 
we introduce at most $O(\frac{n}{s})$ error from linear approximation. 

The other approximations use integrals to approximate their Riemann sums \cite{apostol91,stewart12}:
\begin{equation}
\sum_{i = 0}^{s - 1} \frac{1}{s} f\left(\frac{i}{s}\right) \approx 
\int_{x = 0}^1 f(x) \mathrm{d}x = 1,
\end{equation}
and
\begin{equation}
\sum_{i = 0}^{s - 1} \frac{1}{s} f\left(\frac{i}{s}\right) \ln f\left(\frac{i}{s}\right) \approx
\int_{x = 0}^1 f(x) \ln f(x) \mathrm{d}x. \label{second_integral}
\end{equation}
Since $f$ is monotonically decreasing for $\Bd(1, n)$, the left and right Riemann sums bound the first integral:
\be
\sum_{i = 0}^{s - 1} \frac{1}{s} f\left(\frac{i}{s}\right) \geq
\int_{x = 0}^1 f(x) \mathrm{d}x \geq \sum_{i = 1}^{s} \frac{1}{s} f\left(\frac{i}{s}\right).
\label{eq:riemann-integral-bound}
\ee
The difference between the sums in \eqref{eq:riemann-integral-bound} bounds the difference between
our sum (on the left) and the integral. 
But the difference between the sums is just the terms for $i = 0$ and $s$:
\be
\frac{1}{s} [f(0) - f(1)] = \frac{1}{s} [F'(1) - F'(0)] = \frac{n}{s},
\ee
which introduces $O(\frac{n}{s})$ error.

Following a similar procedure for the other integral (Expression \ref{second_integral}),
\be
\frac{1}{s} [f(0) \ln f(0) - f(1) \ln f(1)] = \frac{1}{s} [n \ln n] = \frac{n \ln n}{s}.
\ee
And that is $O(\frac{n \ln n}{s})$ error.
\end{proof}

\paragraph*{Proof of Theorem \ref{hkn_err_thm}}

\begin{proof}
Following the logic from Theorem \ref{hbs_err_thm}, the expectations over the $\Bd(k, n - k + 1)$ distribution in Theorem \ref{hkn_thm} can be seen as integral approximations to Riemann sums with $O(s)$ terms. 
A Riemann sum that uses the maximum function value over each $\frac{1}{s}$ region in $[0, 1]$ as its term values is an upper bound, and one that uses the minimums is a lower bound, for the Riemann sums we estimate, which use right-edge values for each region, and for the estimating integrals. So the difference between the maximum-value Riemann sum and the minimum-value one bounds the difference between our estimating integrals and the sum they estimate. 

Since our Riemann sums (and the pdf of $\Bd(k, n - k + 1)$) are neither monotonically increasing nor decreasing, 
but instead have a single local maximum in the interior, there may be up to two terms that are different between 
the minimum and maximum Riemann sums. 
So the error is bounded by twice the difference between the greatest term size over the sum and 
the edge values, which are zero. 
The greatest terms are $O(\frac{n}{s})$ for $p_j$. 
Likewise, the pdf of $\Bd(k, n - k + 1)$ at its mode, $\frac{k - 1}{n - 1}$, is an $O(n)$ quantity. 
When we multiply $p_j$ by logarithms in some expressions, the greatest terms are $O(\frac{n \ln n}{s})$.

Regarding other approximations: 
The error in $\psi(z) \approx \ln z$ is $O(\frac{1}{z})$ \cite{digamma_appx}.
When we drop an additive $-1$ term, it is later amortized over $k - 1$, introducing an error of at 
most $\frac{1}{k - 1}$ bits per hash code in the sketch. 
The error in the approximation $\left(1 + \frac{c}{z}\right)^z \approx e^c$ is $O(\frac{1}{z})$. 

Next, there are the approximations for factorials.  
From Expression \ref{from_b} to Expression \ref{to_b} we use Stirling's approximation 
\cite{stirling30,demoivre33,feller68,hald07,wiki_stirling}: 
$\ln(n!) \approx n \ln n - n$. 
That has $O(\ln n)$ error. 
But those errors have some cancellation over the three factorials from Expression \ref{from_b}. 
Using a tighter version of Stirling's approximation: 
$\ln(n!) \approx n \ln n - n + \frac{1}{2} \ln (2 \pi n) + O(\frac{1}{n})$, 
adds the following terms to \eqref{to_b} which are easily bounded as:
\begin{align}
    \eqref{from_b} &\approx \eqref{to_b} -\frac{1}{2} \ln (2 \pi n) + \frac{1}{2} \ln (2 \pi (n - k)) + \frac{1}{2} \ln (2 \pi k) \nonumber \\     
    &= \eqref{to_b} +  \frac{1}{2} \ln \left( 2 \pi k \left(\frac{n - k}{n}\right)\right) \nonumber \\ 
    &\le \eqref{to_b} + \frac{1}{2} \ln \left( 2 \pi k\right)
\end{align}
So the error is $O(\ln k)$ for the whole data sketch, which is $O(\frac{\ln k}{k})$ bits per 
hash code for the $k$ hash codes in the data sketch.
\end{proof}

\section{Evaluating Approximations via Empirical Means} 
\label{emp_mean_sec}

To empirically evaluate the accuracy of the approximation for entropy $H_{K|N}$ from 
Theorem \ref{hkn_thm}, in this section we compare it to empirical estimates based on the concept of empirical entropy \cite{cover06}. 

Empirical entropy is a sample-based estimate of entropy for a distribution. 
The empirical entropy for a sample is the average of $- \ln p(x)$ over observations $x$ in a sample. 
For $H_{K|N}$, the distribution is uniform over size-$n$ subsets of a size-$s$ set, and $p(x)$ is $p_j p_k$. 
So to estimate $H_{K|N}$ via empirical entropy, we could take a sample of size-$n$ subsets, and take the average of $- \ln (p_j p_k)$ over 
the size-$n$ subsets in the sample.

Since $p_j p_k$ depends only on $j$, it is the same for all size-$n$ subsets that have the same $k$th least element $j$. 
As a result, recall that in Expression \ref{hkn_one_term}, reiterated below in \eqref{eq:expectation-over-j-again}, 
we showed that we can write $H_{K|N}$ as an expectation over $j$-values:
\begin{equation}
H_{K|N} = - \sum_{j = k}^{s} p_j \ln (p_j p_k). 
\label{eq:expectation-over-j-again} \tag{$\star$}
\end{equation}
So the average of $- \ln (p_j p_k)$ over $j$-values sampled according to the distribution with pmf $p_j$ is an unbiased estimate of $H_{K|N}$. 
(The $j$-value resulting from drawing a size-$n$ subset and taking the $k$th least value has a distribution with pmf $p_j$.) 
We will use this average to estimate $H_{K|N}$.

Based on Expressions \ref{pj_def} and \ref{pk_def}, for $p_j$ and $p_k$ it can be shown that samples for 
$- \ln (p_j p_k)$ take the form 
\begin{align}
    &\sum_{i = 0}^{n - 1} \ln (s - i) - \sum_{i = 0}^{k - 1} \ln (n - i)  \nonumber \\
    &- (n - k) \ln s - (n - k) \ln (1 - t - \frac{n - k - 1}{s}),
\end{align} 
where the first two terms are constant with respect to $j$, and the last two terms depend on the sampled $t$-values
for $t = \frac{j}{s}$.


\begin{figure}
\begin{tikzpicture}

    \definecolor{crimson2143940}{RGB}{214,39,40}
    \definecolor{darkgray176}{RGB}{176,176,176}
    \definecolor{darkorange25512714}{RGB}{255,127,14}
    \definecolor{forestgreen4416044}{RGB}{44,160,44}
    \definecolor{mediumpurple148103189}{RGB}{148,103,189}
    \definecolor{orchid227119194}{RGB}{227,119,194}
    \definecolor{sienna1408675}{RGB}{140,86,75}
    \definecolor{steelblue31119180}{RGB}{31,119,180}

    \definecolor{blublublu}{RGB}{0, 79, 250}
    \definecolor{redredred}{RGB}{194, 2, 2}
    \definecolor{redred}{RGB}{250, 120, 120}
    \definecolor{_red}{RGB}{250, 200, 200}
    
    \begin{axis}[
    width=\axisdefaultwidth, 
    height= 0.75*7.3cm,
    legend cell align={left},
    legend columns=3,
    legend style={fill opacity=0.8, draw opacity=1, text opacity=1, at={(1.025,1.125)}, anchor=east, draw=none},
    tick align=outside,
    tick pos=left,
    x grid style={darkgray176},
    xlabel={\(\displaystyle \log_2n\)},
    xmajorgrids,
    xmin=9.5, xmax=20.5,
    xtick style={color=black},
    y grid style={darkgray176},
    ylabel={Bits per hash code},
    ytick={-0.135, -0.090, -0.045, 0, 0.045, 0.090, 0.135},
    yticklabels={-0.135, -0.090, -0.045, 0, 0.045, 0.090, 0.135},
    ymajorgrids,
    ymin=-0.163352080626581, ymax=0.147443472516093,
    ytick style={color=black},
    yticklabel style={
        rotate=45, xshift=10pt,
        yshift=-2.5pt,
        /pgf/number format/fixed,
        /pgf/number format/precision=3
    },
    ylabel style = {
        xshift= -5pt
    },
    tick label style={font=\large}, 
    label style={font=\large}, 
    ]
    \addplot [ultra thick, blublublu, opacity=1, densely dotted]
    table {%
    10 0.00476940860676223
    11 -0.0321176871791626
    12 -0.0378746414749205
    13 -0.0415264873019322
    14 -0.0444653045516095
    15 -0.0436780926130415
    16 -0.0459228890240429
    17 -0.0443854450701204
    18 -0.0441604304206346
    19 -0.043921232227705
    20 -0.0478238433142074
    };
    \addlegendentry{$q_1: 0.159$}
    \addplot [ultra thick, blublublu, opacity=0.5, dash dot]
    table {%
    10 0.00476940860676223
    11 -0.0666933663753162
    12 -0.0785816135457209
    13 -0.0857976377932261
    14 -0.08809954486678
    15 -0.0876707812216966
    16 -0.0932933525421395
    17 -0.0890789053409235
    18 -0.0892893406358792
    19 -0.0906940621471464
    20 -0.0907949177563811
    };
    \addlegendentry{$q_2: 0.023$}
    \addplot [ultra thick, blublublu, opacity=0.25,  loosely dashed]
    table {%
    10 0.00476940860676223
    11 -0.106748837043381
    12 -0.113662689645564
    13 -0.126709658257613
    14 -0.13038352266556
    15 -0.131097920464023
    16 -0.140419086128518
    17 -0.149225010029187
    18 -0.145163327243803
    19 -0.134270323743393
    20 -0.138043123216413
    };
    \addlegendentry{$q_3: 0.001$}
    \addplot [ultra thick, redredred, opacity=1, densely dotted]
    table {%
    10 0.00476940860676223
    11 0.0316907014991745
    12 0.0404391063361541
    13 0.042068938166395
    14 0.0444132647037086
    15 0.0436612196141868
    16 0.0454776520173921
    17 0.0444130045191138
    18 0.0460106689762277
    19 0.0441730530903982
    20 0.0450126321283265
    };
    \addlegendentry{$Q_1: 0.841$}
    \addplot [ultra thick, redredred, opacity=0.5, dash dot]
    table {%
    10 0.00476940860676223
    11 0.0619546907861318
    12 0.0759323647034577
    13 0.0827764665124769
    14 0.0861370001652266
    15 0.0873694538571416
    16 0.0884120083678833
    17 0.0894349759930772
    18 0.0912489471773642
    19 0.0875174421415431
    20 0.0875846361348551
    };
    \addlegendentry{$Q_2: 0.977$}
    \addplot [ultra thick, redredred, opacity=0.25,  loosely dashed]
    table {%
    10 0.00476940860676223
    11 0.0897921310243256
    12 0.109265732876019
    13 0.118209227405887
    14 0.123404117253582
    15 0.127285092983256
    16 0.131470070235076
    17 0.122094440292963
    18 0.127166480214381
    19 0.133316401918698
    20 0.130334053109945
    };
    \addlegendentry{$Q_3: 0.999$}
    \addplot [ultra thick, black ]
    table {
    10 0.00476940860676223
    11 -0.000268838207107153
    12 0.000741034081700413
    13 0.000121065485027404
    14 0.000496568356003988
    15 0.000107514962220682
    16 -0.000663643513822852
    17 -5.17778272847721e-05
    18 0.00119261661200603
    19 7.46802371053235e-05
    20 -0.00109314341445394
    };\label{mean}
    \end{axis}
    
    \end{tikzpicture}
    \caption{Mean error  (\ref{mean}) and quantiles of error from estimation with $b = 64$.
    Confidence interval for ``$t$-sigma rule" can be found by evaluating $Q_t - q_t$.
    Figure \ref{gap_plot}  zooms in on mean error curve (\ref{mean}) for different values of $b$.}
    \label{fig:gap_full}
\end{figure}
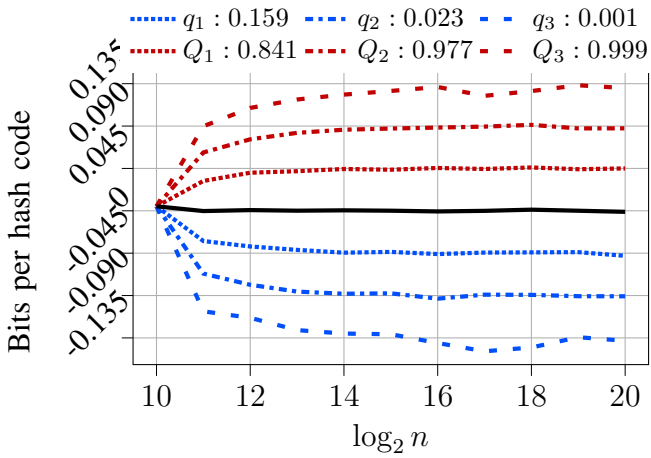
    
\begin{figure}
\pgfplotsset{scaled y ticks=false}
\begin{tikzpicture}

    \definecolor{crimson2143940}{RGB}{214,39,40}
    \definecolor{darkgray176}{RGB}{176,176,176}
    \definecolor{darkorange25512714}{RGB}{255,127,14}
    \definecolor{forestgreen4416044}{RGB}{44,160,44}
    \definecolor{lightgray204}{RGB}{204,204,204}
    \definecolor{steelblue31119180}{RGB}{31,119,180}
    
    \begin{axis}[
        width=\axisdefaultwidth, 
        height= 0.75*7.3cm,
        legend cell align={left},
    legend columns=2,
    legend style={fill opacity=0.8, draw opacity=1, text opacity=1, at={(0.8,1.125)}, anchor=east, draw=none},
    tick align=outside,
    tick pos=left,
    x grid style={darkgray176},
    xlabel={\(\displaystyle \log_2n\)},
    xmajorgrids,
    xmin=9.5, xmax=20.5,
    xtick style={color=black},
    y grid style={darkgray176},
    ylabel={Bits per hash code},
    ytick = {-0.001, 0, 0.001, 0.002, 0.003, 0.004, 0.005},
    ymajorgrids,
    ymin=-0.00147000613225667, ymax=0.00506652359433456,
    ytick style={color=black},
    yticklabel style={
        rotate=45, xshift=5pt,
        /pgf/number format/fixed,
        /pgf/number format/precision=3
    },
    ylabel style = {
        xshift= -5pt
    },
    tick label style={font=\large}, 
    label style={font=\large}, 
    ]
    \addplot [very thick, steelblue31119180, mark=square*, mark size=1, mark options={solid}]
    table {%
    10 0.00476940860676223
    11 -0.000268838207107153
    12 0.000741034081700413
    13 0.000121065485027404
    14 0.000496568356003988
    15 0.000107514962220682
    16 -0.000663643513822852
    17 -5.17778272847721e-05
    18 0.00119261661200603
    19 7.46802371053235e-05
    20 -0.00109314341445394
    };
    \addlegendentry{$b=64$}
    \addplot [very thick, darkorange25512714, mark=asterisk, mark size=1, mark options={solid}]
    table {%
    10 0.00476940860676223
    11 -0.000425008578365424
    12 -0.000563512485847167
    13 -0.000660962709948524
    14 0.000620436716727207
    15 -3.47348496296007e-05
    16 -8.88653895554057e-05
    17 0.000859319180673049
    18 -0.000278907005164299
    19 -0.000555880084957633
    20 0.000376288819648232
    };
    \addlegendentry{$b=128$}
    \addplot [very thick, forestgreen4416044, mark=diamond*, mark size=1, mark options={solid}]
    table {%
    10 0.00476940860676223
    11 0.00014861698723596
    12 -0.000993318011619147
    13 -0.00103855561467506
    14 -0.000312260961012566
    15 -1.80878043595044e-05
    16 0.000256261064494788
    17 0.00141260954694922
    18 0.000376521703861096
    19 -0.000560797556902446
    20 0.000364280896150188
    };
    \addlegendentry{$b=256$}
    \addplot [very thick, crimson2143940, mark=*, mark size=1, mark options={solid}]
    table {%
    10 0.00476940860676223
    11 -0.000665550857393055
    12 -0.00117289114468434
    13 0.000626308084131483
    14 -0.00011033963167913
    15 0.000674131538363156
    16 -8.75154124200208e-05
    17 0.000274467282624221
    18 0.000201661340359578
    19 -0.000433599628463583
    20 0.000465638488527088
    };
    \addlegendentry{$b=512$}
    \end{axis}
    
\end{tikzpicture}
    
\caption{Comparison of sample-based estimate of $H_{K|N}$ to the approximation from Theorem \ref{hkn_thm}, 
with $k = 1024$. 
The sample-based estimates are based on $5000$ samples each. 
The small differences indicate that Theorem \ref{hkn_thm} is an accurate approximation.
In general, the differences decrease with increasing hash key length. 
}
\label{gap_plot}
\end{figure}
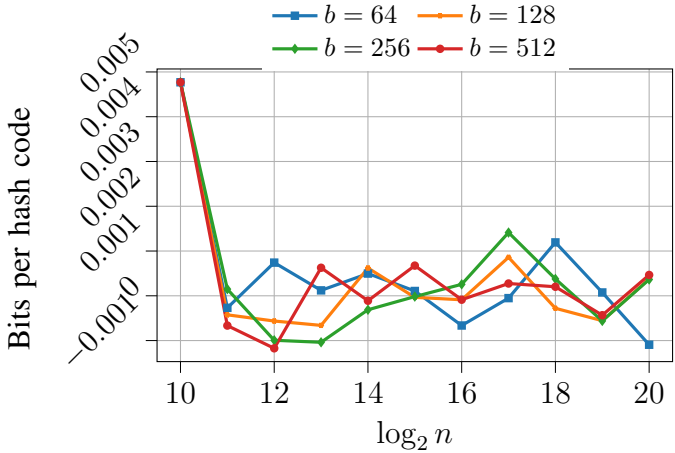

Figures \ref{fig:gap_full} and \ref{gap_plot} show differences between the approximate
lower bound from Theorem \ref{hkn_thm} and the empirical entropy sample average for $H_{K|N}$ over samples of $t$. 
Plotted on the $y$-axis are \emph{differences} between the lower bound and the sample average which concentrate about $0$, 
showing that our sample averages is often close to the lower bound.
The differences are in terms of bits per hash code, so the numbers for $H_{K|N}$, divided by $k$, and converted to base 2 logarithms. 
In absolute terms (before differences), the sample averages are several orders of magnitude greater than the sample standard deviations, indicating high accuracy with high confidence. 
For example, for $s = 2^{64}$, $n = 2^{16}$, and $k = 2^{10}$, in bits per hash code, the sample average is about 49, and the sample standard deviation is about 0.045, 
reflected in Figure \ref{fig:gap_full}.
Increasing the length of the hash code decreases noise in the sample mean, 
as illustrated in Figure \ref{gap_plot}.

\end{appendices}
\end{document}